# High-Intensity and Uniform Red-Green-Blue Triple Reflective Bands Achieved by Rationally-Designed Ultrathin Heterostructure Photonic Crystals

Lu Qiu[1,2#], Quan-Shan Liu[1,2#], Rui Zhang[1,2*], Tao Wen[1,2*]

[1]South China Advanced Institute for Soft Matter Science and Technology, School of Molecular Science and Engineering, South China University of Technology, Guangzhou 510640, China.

[2]Guangdong Provincial Key Laboratory of Functional and Intelligent Hybrid Materials and Devices, South China University of Technology, Guangzhou 510640, China.

**ABSTRACT**:

The relationships between material constructions and reflective spectrum patterns are important properties of photonic crystals. One particular interesting reflectance profile is a high-intensity and uniform three-peak pattern with peak positions right located at the red, green, and blue (RGB, three original colors) region. For ease of construction, a seek for using one-dimensional photonic crystals to achieve RGB triple reflective bands is a meaningful endeavor. Only very limited previous studies exist, all relying on traditional periodic photonic crystals (PPCs) and of large thickness. The underlying physical principles remain elusive, leaving the question of thickness limit to achieve RGB bands unaddressed. Here, we present the first detailed work to explore the thickness limit issue based on both theoretical and experimental investigation. A set of heuristically derived design principles are used to uncover that the break of translational symmetry, thus introducing heterostructure photonic crystals (HPCs), is essential to reduce the total optical path difference (OPD) to ~ 3200nm (the theoretical limit) while still exhibiting high-quality RGB bands. A systematic experiment based on a 12-layer heterostructure construction was performed and well confirmed the theoretical predictions. The associated three-peak properties are successfully used to realize quantum dot fluorescent enhancement phenomena. Furthermore, the HPC exhibits unusually stability against solvent stimulus, in strong contrast to typical behaviors reported in PPCs. Our work for the first time proposes and verifies important rational rules for designing ultrathin HPCs toward RGB reflective bands, and provides insights for a wider range of explorations of light manipulation in photonic crystals.

Photonic crystals (PCs) represent a class of periodic optical materials composed of alternating media with different dielectric constants.[1,2] Interference of light reflected at the interfaces between high- and low-refractive-index domains gives rise to a photonic band gap, at which strong reflectance results in remarkable structural coloration.[1–3] Taking advantage of simple configurations and convenience of fabrication, one-dimensional (1D) PCs are widely studied and applied in various areas, such as sensor technology,[4–8] integrated sensing platforms,[9,10] and solar cells.[11] White light that mimics natural daylight can be created using individual red, green, and blue light sources, *e.g.*, light-emitting diodes (LEDs), and the color variability can be achieved by tuning the desired color point of the lamp.[12–14] One of the most important applications of 1DPCs is to enhance brightness without consuming more power.[15] 1DPCs simultaneously possessing three band gaps corresponding to the wavelength of three primary colors are desired by white-light illumination. However, up to now examples of 1DPCs for multicolored/white light illumination have been rarely reported. Existed attempts all depended on large-thickness 1D periodic PCs (PPCs) stacked by the same repeating units composed of alternating dielectric media.[16,17] For both scientific curiosity and practical fabrication benefit, it is interesting to ask what the 1DPC thickness limit is in order to achieve high-intensity and uniform RGB triple reflective bands. In this work, we heuristically derived a set of design principles underlying the reflective peak positions, amplitudes, and uniformity in the visible spectrum. To the best of our knowledge, these rules have not been summarized in previous studies around the relationships between material constructions and reflective spectrum patterns of 1DPCs. Applying the rules allows us to convincingly uncover, for the first time, that breaking translational symmetry of PPCs by introducing heterostructure PC (HPC) construction is necessary to reduce the total optical path difference (OPD) to around 3200nm, the theoretical minimal value, to still achieve good-quality RGB triple reflective bands.

HPC systems have already been used in some application scenarios, where a breaking of translational symmetry due to the absence of rigorous periodicity as in PPCs provides an essential mechanism basis to generate desired advantageous properties. The most common feature of HPC architectures is band gap-broadening effect. Notable examples include enlargement of omnidirectional total reflection frequency range in 1DPC,[18] enlargement of the nontransmission frequency range of 1D multiple-channeled filters,[19] achievement of universal fluorescence enhancement for commonly used fluorescent media on a multiple HPC with a super-wide stopband,[20] realization of optical broadband angular selectivity by combining multiple 1DPCs with different periodicities.[21] However, the aforementioned HPC systems are only applicable to circumstances asking for broad spectral ranges of up to hundreds of nanometers. For specifically needed spectral patterns such as multiple reflective bands at distinct specified locations, broadened band gaps become powerless to meet the requirement. Our work represents another remarkable contribution toward rational control of light reflectance patterns based on the HPC idea. We have systematically investigated this new ultrathin HPC concept based on a coupled theoretical-experimental study. Construction of 1D HPC possessing high reflective bands (HRBs) at RGB regions simultaneously following theoretical design principles was realized experimentally, which exhibited RGB triple reflective bands in good agreement with theoretical predictions. It is demonstrated that the fluorescence emission of red/green/blue quantum dots can be enhanced by the corresponding reflection of this experimental 1D HPC. Moreover, the response of it to organic solvent vapor was examined, and an unique dynamic behavior was observed.

To begin, we first analyze 1D PPCs (also known as distributed Bragg reflectors[22,23]) with single structural periodicity, which give major reflection peaks at certain

wavelengths at normal incidence, determined by the Bragg-Snell law:[22–24]

$$\lambda_{peak} = 2(n_L d_L + n_H d_H)/a \quad (1)$$

where $n_H$ and $n_L$ are the refractive indices of the high- and low-index layers and $d_H$ and $d_L$ are the thicknesses of the high- and low-refractive-index layers, respectively, and $a$ is a positive integer number representing the diffraction order. Strictly speaking, additional mathematical constraints are required on reflective indexes/thicknesses to avoid internal destructive interference that destroys the appearance of particular peaks, but Eq. 1 suffices for the discussion related to the theme of this work. Because a RGB triple reflective bands corresponds to the simultaneous appearance of reflectivity peaks at three wavelengths in the red, green, and blue region, respectively, Eq. 1 can be used to deduce a realization strategy based on PPC systems by the following equation:

$$\lambda_{peak,red}: \lambda_{peak,green}: \lambda_{peak,blue} = 1/a_1: 1/: a_2: 1/a_3 \quad (2)$$

Substituting into specific wavelength values corresponding to the three primary colors, the minimal integer solution of $(1/a_1, 1/, a_2, 1/a_3)$ can be determined as $(1/5, 1/6, 1/7)$. The resulting minimal optical thickness of $n_L d_L + n_H d_H$ is thus 1550nm-1650nm, which leads to RGB triple peaks around 620 to 660 nm, 517 to 550 nm, and 443 to 471 nm, respectively. While not explicitly articulated the above mathematical analysis, previous reports of RGB triple reflective bands essentially all adopted this strategy to construct the repeating unit (also called a stack) in a PPC fashion. However, to gain high peak intensities, the number of stacks needs to far exceed one (in fact 5 as in ref. 16 and 11 as in ref. 17), significantly increasing the total optical thickness.

To tackle the thickness limit problem in a quantitative fashion, we raised a conjecture beyond Eq. 1, which constitutes the theoretical basis to reveal the detailed difference between 1D PPCs and HPCs for the first time, as well as extract a series of design principles toward realizing RGB triple bands in HPC systems at the ultrathin limit. Our conjecture starts from a consideration that can be conveniently thought in an air/dielectric layers/substrate construction where the refractive index contrast at the two major interfaces are much stronger than at the internal interfaces. At normal incidence, the Fresnel reflection coefficient which decides the amplitude of reflectivity is determined by the refractive index ratio of dielectric media (Supporting Information, section 1). Therefore we primarily consider the interference effect of light reflected from two material interfaces involving air and substrate, respectively. The optical path difference (OPD) between the two specified beams of reflected light (Figure 1b) is twice the total optical thickness of the 1DPC. In other words, the OPD can be written as $2 \times \sum n_i d_i$ (here the subscript $i$ may indicate any individual layer). Constructive interference of the two beams of reflected light similar to Eq. 1 is conjectured to occur when

$$a \times \lambda = \text{OPD} = 2 \times \sum n_i d_i \tag{3}$$

where $a$ is a positive integer and $\lambda$ is the corresponding reflective band position. Based on Eq. 2, we can deduce that the minimal OPD would be 3100nm-3300nm in order to realize three peaks in RGB regions respectively. The situation is totally different now in that multiple dielectric layers can be designed. In contrast, only two dielectric layers exist in Eq. 1. To make a concrete connection to real materials, we introduced in our model and experimental set-up that include titania sol (TiO$_2$) and UV-curable resin (resin) which were chosen to be the dielectric media, while air and silicon (Si) respectively served as two semi-infinite ambient media. Their refractive index values are $n_1$ (TiO$_2$) = 1.80, $n_2$ (resin) = 1.55, $n_{Air}$ = 1.00, and $n_{Si}$ = 4.33. Our choices of dielectric layers and substrate are built on our previous work and specific fabrication techniques developed in our laboratory.[25] We note that the general design principles proposed in this work are not tied to any specific material parameters. In order to be consistent on theoretical calculations and experimental tests, we focus on these materials to demonstrate the design principles of RGB triple bands in all theoretical calculations presented in the main text. The inclusion of resin provides additional flexibility for our system to study stimuli-responsive behaviors. In the case of 1D PPC, it is unable to realize uniform high-intensity RGB triple peaks based on Eq. 3 and more than one repeating unit. One simple argument is that the major peaks derived by Eq. 1 always dominate, but it is impossible to emerge RGB triple major peaks for more than one repeating unit and OPD~3200nm.

We envision that properly designed 1D HPC systems could create the possibility to maintain total OPD~3200nm and by carefully adjusting internal constructive and destructive interference to achieve good-quality RGB triple peaks. To demonstrate this novel design idea in some concrete realizations, we consider a generic class of HPC system as Figure 1a shows its schematic construction, which is a heterostructure composed of two different 1DPCs. Figure 1b shows a concrete system construction used in our work to test this idea. Here, the two 1DPCs share the same material composition (again $TiO_2$ and resin in our detailed calculation) and layer number, but with different layer thicknesses. *n* and *d* respectively represent the refractive index and the thickness of any individual layer. The 2*m*-layer heterostructure then can be represented as Air/*m*-layer thinner 1DPC/*m*-layer thicker 1DPC/Si substrate. To maintain the integrity of the 1DPC structure, *m* has to be some positive even number. To further reduce the parameter space while still capturing the basic design scheme, the resin thicknesses in eachstack are identical. Three instead of four independent layer thicknesses additionally mitigated the challenge of experimental fabrication.

We first perform some preliminary analysis based on the scheme shown in Figure 1. Our goal now is to explore possible HPC constructions that follow the design of Figure 1b, possess OPD~3200nm, and present high-intensity and uniform RGB triple peaks. We define the OPD for a stack (an alternating high-index and low-index dielectric layer) as partial OPD (p-OPD). There are only two different kinds of stacks in all 2*m*-layer heterostructures. Consequently, there are only two different types of partial OPDs, represented by $2(n_1 d_1 + n_2 d_2)$ and $2(n_1 d_3 + n_2 d_4)$ respectively. The central speculation is that if the p-OPDs of the stacks in two 1DPCs are reasonably different but both have a magnitude comparable to RGB wavelengths, then it is possible to generate good-quality RGB triple peaks based on cooperative internal constructive and destructive interference due to the breaking of translational symmetry. Considering four different *m* values: 2, 4, 6, and 8, the corresponding average p-OPD values are thus 1600nm, 800nm, 533nm, and 400nm, respectively, with only for m=6 the average value is in the RGB region. We can therefore derive mathematically that for m=2, 4, 8, and beyond, at least one p-OPD has to locate outside the RGB region. Our detailed analysis shows that the Figure 1a speculation indeed works for m=6 in some reasonable thickness value combinations. Interestingly, we also found that for m =2 and 4, good-quality RGB triple peaks can form given one p-OPD must in the RGB region. Below we present more details about these calculations and our novel understanding gained thusly motivated by the Figure 1a idea.

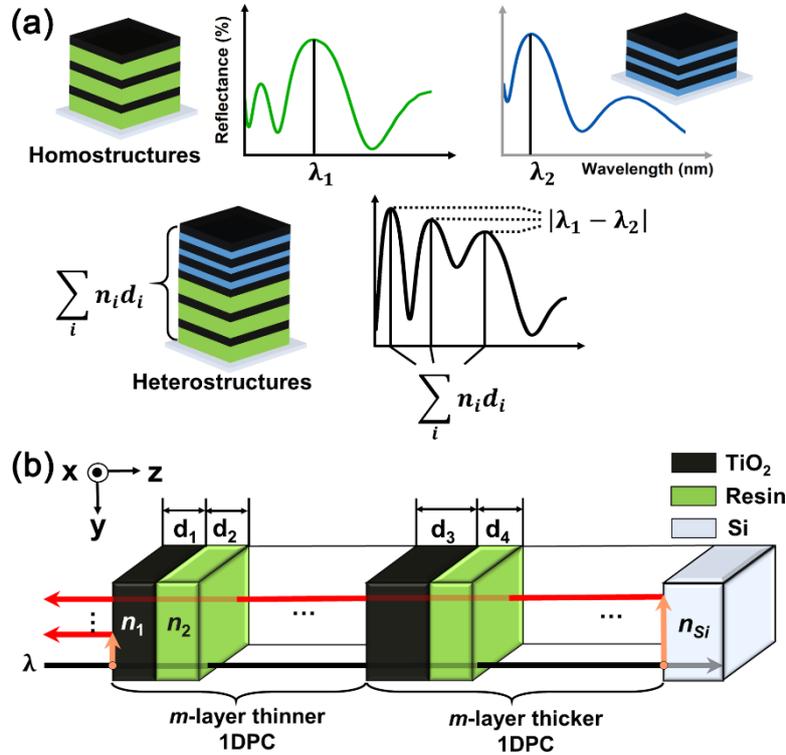

**Figure 1.** (a) The schematic of the design strategy of the 1-D photonic heterostructure. (b) The schematic construction of a 2*m*-layer heterostructure, and the optical path difference between two specified beams of reflected light. The lower black line represents incident light while the upper red lines both refer to reflected light. The orange lines are set to distinguish the reflected light, and they are not the actual light path. In the calculations and experiments discussed in this work, we fix $d_2=d_4$.

In Figure 2, theoretical simulations were conducted to present the multiple reflective bands phenomenon. In Figure 2a (m=2) and 2b (m=4), all groups of material thicknesses are chosen to assure that one kind of stack could bear a partial OPD located near RGB realms. It should be noted that choices of material thickness combinations can be numerous, and thus only several of them were picked out to illustrate the multiple reflective bands effect. On the other hand, the adoption of different thickness combinations also conduces to verifying the applicability of the OPD-based guideline.

For m=2, we designed 4 sets of material thicknesses, noted as 4-1, 4-2, 4-3, and 4-4 respectively. For convenience, we represent precise thickness values in the formulation of $d_1|d_2|d_3|d_4$ (/nm). Therefore, we have 50|110|660|110 for 4-1 (OPD = 3238 nm), 70|85|680|85 for 4-2 (OPD = 3227 nm), 90|60|700|60 for 4-3 (OPD = 3216 nm), and 110|35|720|35 for 4-4 (OPD = 3205 nm). The simulations were conducted at normal incidence using Rouard's method[26] throughout this article and the wavelength ranged from 400 to 800 nm. It can be observed that three reflective bands emerge at RGB regions in all listed cases, which indicates that the simulation results highly agree with the calculation results, thus verifying the credibility of the OPD principle suggested by Eq. 3. Meanwhile, the average intensities of the three reflective bands fluctuate between 45 to 55% and present good uniformity. The trends are similar for $m = 4$, where we designed another 4 sets of material thicknesses for the 8-layer heterostructure, denoted by 8-1, 8-2, 8-3, and 8-4 respectively (Figure 2b). Regarding precise thickness values, we have 95|110|170|110 for 8-1 (OPD = 3272 nm), 110|85|195|85 for 8-2 (OPD = 3250 nm), 120|60|220|60 for 8-3 (OPD = 3192 nm), and 135|35|245|35 for 8-4 (OPD = 3170 nm). It can be seen that three reflective bands are well located at RGB regions in all listed cases. Regarding the average intensities of the three reflective bands, the 8-layer heterostructure has a similar performance (~45% to 55%) compared to those of the 4-layer heterostructure. It is noticeable that for both m=4 and m=8 cases, there might be appreciable intensity gaps among 4 sets of reflection spectra. We suppose it can be attributed to overall optical interference affected by multiple reflections between adjacent layers. If necessary, further optimization can be performed using a brute-force approach which utilizes a computer to automatically search out thickness combinations space segmented by carefully designed thickness intervals. We also studied the scenario where both two categories of stacks hold their partial OPDs beyond RGB ranges, and as expected the peak values decrease noticeably (Figure S1a and b).

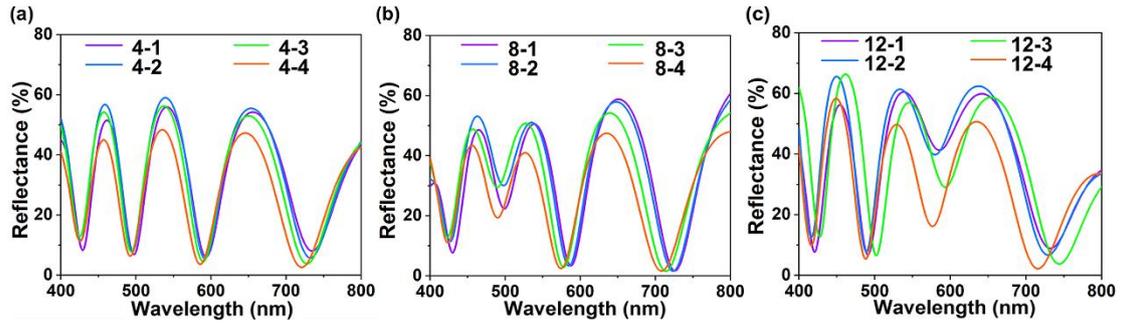

**Figure 2.** (a) The reflection spectra of 4-layer heterostructures with 4 groups of material thicknesses. (b) The reflection spectra of 8-layer heterostructures with 4 groups of material thicknesses. (c) The reflection spectra of 12-layer heterostructures with 4 groups of material thicknesses.

For $m = 6$ which is an aforementioned special case, 4 sets of material thicknesses were adopted for the 12-layer heterostructure, denoted by 12-1, 12-2, 12-3, and 12-4 respectively (Figure 2c). Regarding precise thickness values, we have 35|110|70|110 for 12-1 (OPD = 3180 nm), 50|85|95|85 for 12-2 (OPD = 3147 nm), 70|60|125|60 for 12-3 (OPD = 3222 nm), and 85|35|145|35 for 12-4 (OPD = 3135 nm). It can be observed that three reflective bands emerge at RGB regions in all listed cases, and their average intensities (~55 to 60%) are greater than those of the 4- and the 8-layer heterostructure. By carefully designing material thicknesses, both two sorts of stacks can hold partial OPDs inside the visible light spectrum, and hence the three band intensities of the 12-layer heterostructures have a better performance than those of the 4- and 8-layer heterostructures. It should be noted that the two partial OPDs also reveal the stop-band positions (SBPs) of the two constituent 1DPCs in the 12-layer heterostructure, since the two partial OPDs are equal to the two SBPs in magnitude respectively. To graphically present the magnitude difference of the two partial OPDs, we could illustrate the relative distance between the two SBPs by calculating the reflection spectra of the two constituent 1DPCs. As further discussed in the Supporting Information, the magnitude difference of the two partial OPDs should be well controlled to make a good uniformity of the three band intensities for m=6.

Furthermore, we can consider the circumstance of the 16-layer heterostructure.

There are 8 stacks and each stack has to bear 400 nm averagely. It can be predicted that the three band intensities of the 16-layer heterostructure won't be greater than those of the 12-layer heterostructure, for the same reason mentioned in the 4- and 8-layer heterostructures case. By limiting the total OPD in 3100 to 3300 nm, we also tried a range of material thicknesses for the 16-layer heterostructure in simulation, but we couldn't even acquire good uniformity of the three band intensities (Figure S1c). Moreover, the same analysis process can be applied for the higher-layer heterostructure, and predictably the three band intensities won't be better.

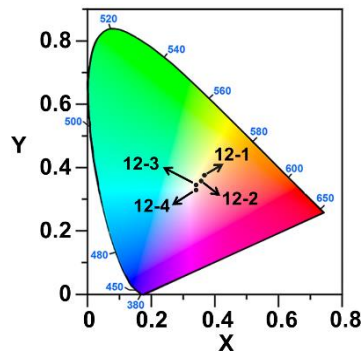

**Figure 3**. CIE 1931 chromaticity coordinates of 12-layer heterostructures with 4 groups of material thicknesses.

In subsequent contexts, we focus on the 12-layer heterostructure which has a relatively better performance on band intensities. With the reflection spectra in Figure 2c, corresponding CIE coordinates were obtained (Figure 3). It is indicated that all the coordinates lie at a reddish-white region, close to the boundary line of the green and red color domain. It can be majorly explained by the asymmetry of the two troughs between the three reflective bands. Though the three reflective bands share a similar intensity, the higher trough at longer wavelengths contributes to the deviation of the CIE coordinates, from the ideal white light point (0.33,0.33) towards the green-red boundary line. To enhance the purity of white light reflection, the simplest solution is to adjust those CIE coordinates to the green-red boundary line. This can be realized by tuning the red light reflection caused by the 5$^{th}$ order reflective band. Considering that a change of the total OPD magnitude has more impact on the 5$^{th}$ than 6$^{th}$ and 7$^{th}$ order reflective bands, we could properly reduce the total OPD to alleviate the red light reflection. Specifically, we set material thicknesses as 50|75|88|75 respectively. As a result, the OPD is decreased to approximately 2885 nm. Therefore, the 5$^{th}$, 6$^{th}$, and 7$^{th}$ order reflective bands are supposed to emerge at about 577 nm, 481 nm, and 412 nm respectively. The theoretical reflection spectrum and the CIE coordinate are demonstrated in Figure 4. It can be observed that three high reflective bands appear around the predicted positions and have similar intensities. On the other hand, the modified CIE coordinate (0.35,0.39) is located at the red-green boundary line as desired. The uniformity of the three band intensities shown in Figure 2c is not a universal phenomenon. However, by comparing the relative magnitudes of the two partial OPDs held by two different stacks, it is indicated that the mentioned uniformity is strongly influenced by the magnitude difference of the two partial OPDs (Figure S2).

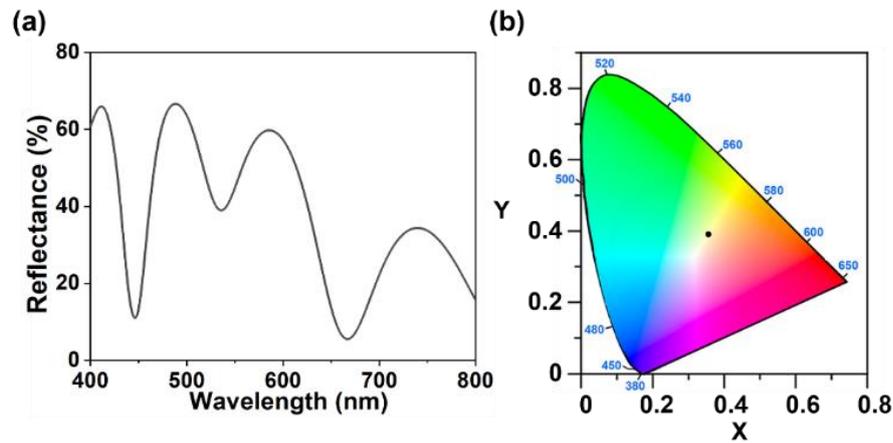

**Figure 4.** (a) The reflection spectrum of the 12-layer heterostructure after decreasing the OPD. (b) CIE 1931 chromaticity coordinate after modification.

Experimentally, a 12-layer heterostructure was fabricated by alternately spin-coating UV-curable resin and titania sol onto a silicon substrate (for the fabrication details see Supporting Information, Section 4), with a cross-sectional SEM image (Figure S3). A thicker 1DPC consisting of 3 stacks was prepared on the substrate firstly, and then a thinner one with the same stacking number was deposited subsequently. The optical properties of the constituent layers were measured by ellipsometry, from which the thicknesses of the polymer layers were determined to be ~100 nm (for both the thick and the thinner 1DPC), while that of the titania layers were ~50 nm (for the thinner

1DPC) and ~88 nm (for the thicker 1DPC), respectively. The experimental layer thicknesses were chosen according to the theoretical thicknesses as suggested above. Here we adopted a semi-empirical thickness scaling relationship, where the ratio between the experimental and the simulated resin thicknesses is approximately 4/3. The scaling relationship can be verified from a series of comparisons between the experimental and the simulated reflection spectra (Figure 5). The origin of this scaling relationship may be attributed to the dispersion of refractive indices mentioned later.

The reflection spectrum of the experimental sample was measured at normal incidence, as shown in Figure 5d. It can be seen that the experimental result is well captured by the simulation result, especially in respect of the three HRBs positions. The simulated reflection spectrum is slightly different from that in Figure 4a for the following reason. Due to the dispersion of light, the refractive indices of the experimental materials may change as the incident light wavelength varies (Figure S4). By considering this effect, a more realistic simulated reflection spectrum was obtained. On the other hand, we ignored light absorption occurring in different media during the simulation process, because the light absorption losses in our experimental materials are relatively low (Figure S5). As for the discrepancies between the measured and the simulated results, there is a range of possible reasons such as material imperfection, weak fluctuations in layer orientation, finite interface width between adjacent layers, and limited collimation of the incident light. To make a full picture of the HRBs behavior, we also recorded the reflection spectra of samples ranging from Air/3-stack thicker 1DPC/Si substrate (short for 0+3) to Air/2-stack thinner 1DPC/3-stack thicker 1DPC/Si substrate (short for 2+3), as shown in Figure 3a-c. Moreover, to investigate the sequence effect caused by 1DPC deposition sequence, we also simulated the reflection spectra of samples ranging from Air/3-stack thinner 1DPC/Si substrate to Air/3-stack thicker 1DPC/3-stack thinner 1DPC/Si substrate (Figure S6). It is indicated that the sequence effect has a limited impact on the positions and the intensities of the three HRBs. However, the two troughs between the three HRBs both experience noticeable intensity fluctuations.

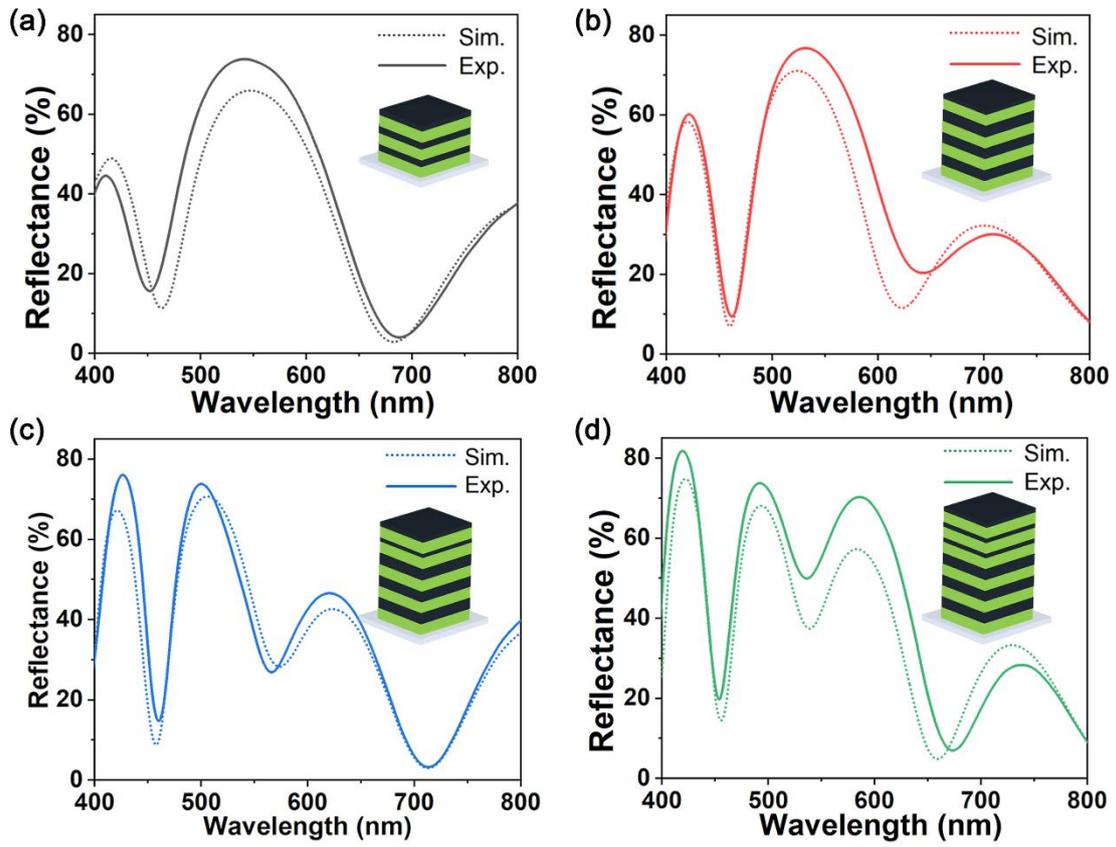

**Figure 5.** Comparison between the simulated (dot line) and the measured (solid line) reflection spectra of samples in different stages (a) 0+3. (b) 1+3. (c) 2+3. (d) 3+3.

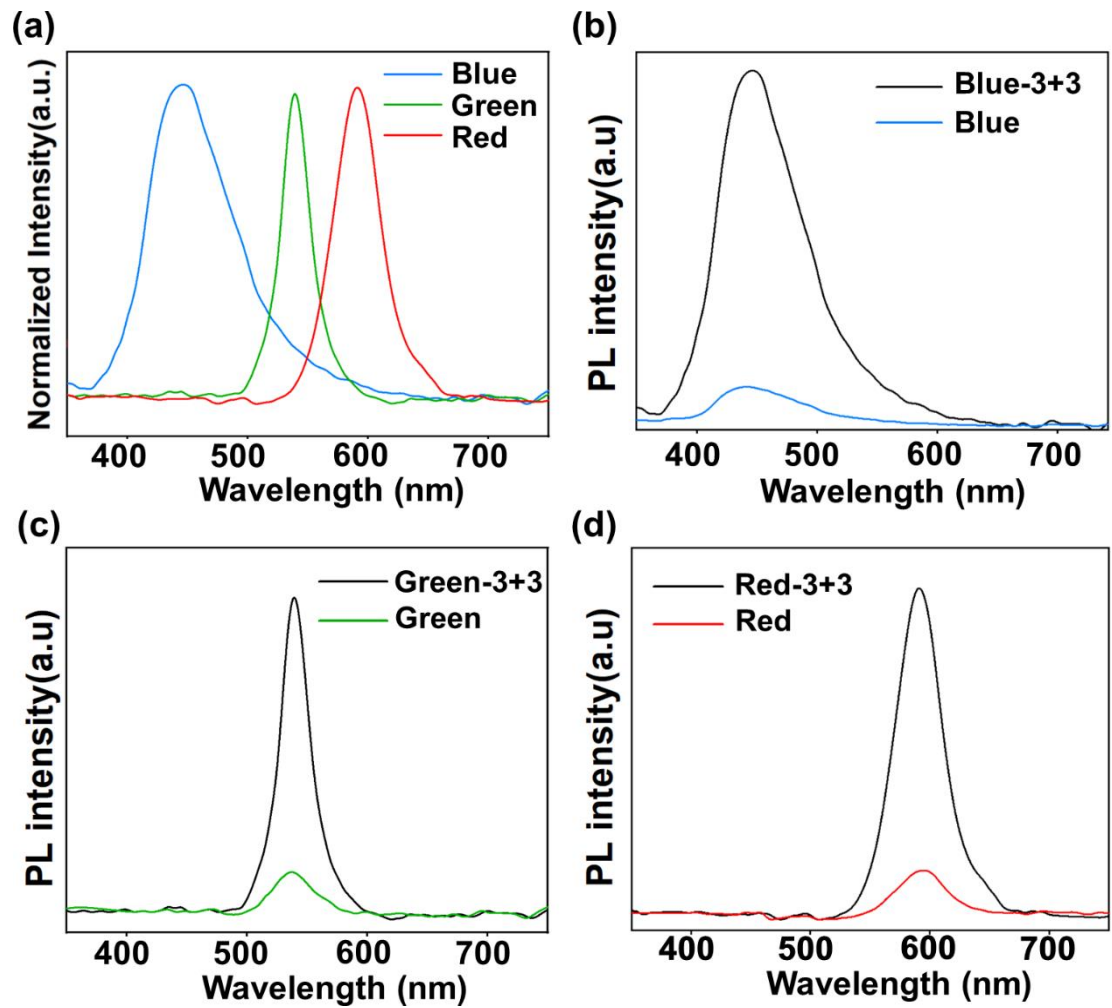

**Figure 6.** (a) the PL spectra of the luminescent materials on bare quartz. (b-d) Enhanced emission of the luminescent materials on the top of the fabricated 12-layer heterostructure.

Furthermore, similar to 1DPCs, the heterostructures in this article can also provide high reflectance at certain wavelength ranges, thus an important application of them is to enhance the collection efficiency of light emission by tuning their HRBs to match with the emission of luminescent materials. To demonstrate the function of enhanced emission, three types of luminescent materials, whose emission peaks are covered in RGB regions respectively, were combined with the experimentally fabricated 12-layer heterostructure. The CdSe (Red, $\lambda$ = 600 nm) / $CH_3NH_3PbBr_3$ (Green, $\lambda$ = 530 nm) / TPE (Blue, $\lambda$ = 430 nm) were respectively coated on a mica substrate covered with carbon film. These luminescent thin films can be floated onto water surfaces and then transferred onto either the heterostructure or bare quartz. The emission spectra of the luminescent layers on bare quartz are shown in Figure 6a, while those of the luminescent layers on the heterostructure are illustrated in Figure 6b-d. It is indicated that the heterostructure notably enhanced the fluorescence of the CdSe by 6.7 times, the $CH_3NH_3PbBr_3$ by 3.1 times, and the TPE by 3.4 times. The above results demonstrate that our heterostructure has well-function in enhancing RGB or white light emission.

Figure 7a shows four pairs of CIE coordinates converted from the four experimental reflection spectra presented in Figure 3. It can be observed that the experimentally fabricated 12-layer heterostructure indeed made a CIE coordinate in the white light region. Besides, the fabricated 12-layer heterostructure is composed of organic and inorganic materials, and the polymer layers appear to be extremely sensitive to solvents, which indicates that swelling may occur when some solvent molecules enter the polymer network. In our follow-up work, dichloromethane was used as the solvent to test the stimulus-response effect of the heterostructure. Under the solvent stimulation, the polymer layers reacted quickly, resulting in an optical thickness change of the heterostructure. As a result, we observed very interesting HRBs' shifts in the optical properties of the heterostructure (Figure 7b). Different from the traditional CIE coordinate shift based on one major reflective band movement of 1DPCs,[25,27–30] the heterostructure in this paper exhibited peculiar spectral behavior when faced with the solvent stimulation. In short, the heterostructure maintained a high reflection effect on RGB emission with the CIE coordinate moving slightly and annularly in the white light region (Figure 7a).

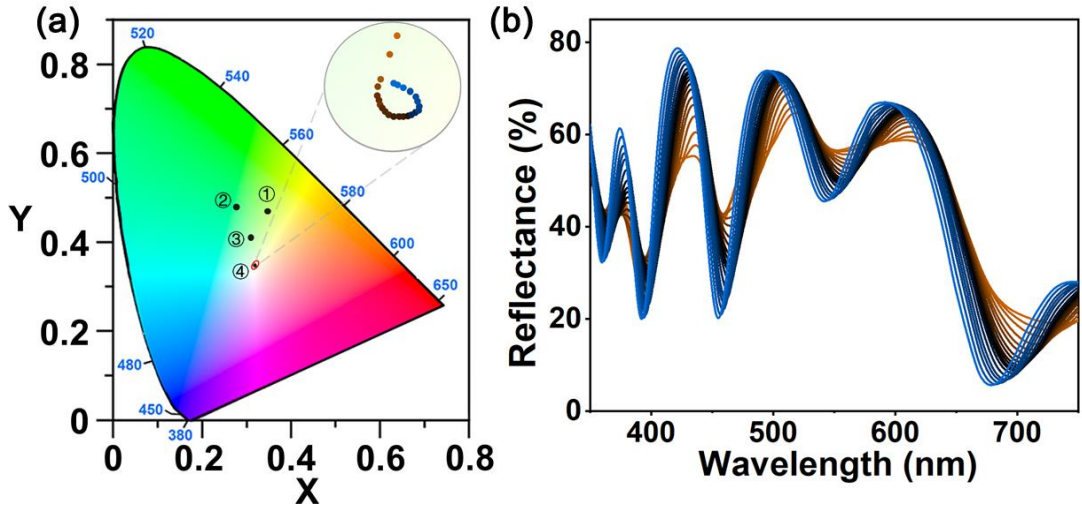

**Figure 5.** (a) CIE 1931 chromaticity coordinates of samples in different stages and the CIE coordinate shift of the fabricated 12-layer heterostructure (0+3 ①; 1+3 ②; 2+3 ③; 3+3 ④) during the stimulation response test. (b) Optical properties of the fabricated 12-layer heterostructure with a response timer.

In summary, through jointly theoretical-experimental efforts, we devised a series of 2m-layer 1DPC-based heterostructures to achieve white light reflection by generating three reflective bands at RGB frequency ranges simultaneously. Then, we investigated how the m value would influence the three reflective bands, and continued our research on the 12-layer heterostructure. Moreover, the positions of the three reflective bands are primarily determined by the total optical thickness of the research system, while the three band intensities are combined results of overall optical interference, multiple reflections between adjacent layers, and the magnitude difference of the two partial OPDs held by two kinds of stacks. In this respect, the three reflective bands can be tuned according to practical needs by carefully choosing dielectric media and controlling material thicknesses. Subsequent studies suggest that our heterostructure could present certain robustness under the solvent stimulation and enhance the fluorescence of QDs. To the best of our knowledge, this is the first report on the 1-D photonic heterostructure with rationally designed simple constructions for RGB reflection. We believe that this work can provide a new insight into the design and application of 1DPCs.

Potential application areas of the heterostructure may include displays, lightings, sensors, and so forth.

**AUTHOR INFORMATION**

Corresponding Author

*E-mail: twen@scut.edu.cn.

*E-mail: rzhang1216@scut.edu.cn.

Author Contributions

#These authors contributed equally to this work.

Notes

The authors declare no competing financial interest.

**ACKNOWLEDGEMENTS**

We acknowledges the support from the National Natural Science Foundation of China (U1832220, 21803020, 51890871, 21973033), the Fundamental Research Funds for the Central Universities (2018ZD13), and Guangzhou Science & Technology Program (201904010345). This work is also supported by the Guangdong Provincial Key Laboratory of Functional and Intelligent Hybrid Materials and Devices (2019B121203003), and the Natural Science Foundation of Guangdong (2016ZT06C322).

**REFERENCE**

(1)   Maldovan, M.; Thomas, E. L. *Periodic Materials and Interference Lithography: For Photonics, Phononics and Mechanics*; John Wiley & Sons, 2009.

(2)   Kinoshita, S.; Yoshioka, S.; Miyazaki, J. Physics of Structural Colors. *Reports*

*Prog. Phys.* **2008**, *71* (7), 76401.

(3) Macleod, H. A. *Thin-Film Optical Filters*; CRC press, 2017.

(4) Goyal, A. K.; Dutta, H. S.; Pal, S. Recent Advances and Progress in Photonic Crystal-Based Gas Sensors. *J. Phys. D. Appl. Phys.* **2017**, *50* (20), 203001.

(5) Inan, H.; Poyraz, M.; Inci, F.; Lifson, M. A.; Baday, M.; Cunningham, B. T.; Demirci, U. Photonic Crystals: Emerging Biosensors and Their Promise for Point-of-Care Applications. *Chem. Soc. Rev.* **2017**, *46* (2), 366–388.

(6) Chan, E. P.; Walish, J. J.; Urbas, A. M.; Thomas, E. L. Mechanochromic Photonic Gels. *Adv. Mater.* **2013**, *25* (29), 3934–3947. https://doi.org/10.1002/adma.201300692.

(7) Lova, P.; Bastianini, C.; Giusto, P.; Patrini, M.; Rizzo, P.; Guerra, G.; Iodice, M.; Soci, C.; Comoretto, D. Label-Free Vapor Selectivity in Poly(p-Phenylene Oxide) Photonic Crystal Sensors. *ACS Appl. Mater. Interfaces* **2016**, *8* (46), 31941–31950. https://doi.org/10.1021/acsami.6b10809.

(8) Zhang, R.; Wang, Q.; Zheng, X. Flexible Mechanochromic Photonic Crystals: Routes to Visual Sensors and Their Mechanical Properties. *J. Mater. Chem. C* **2018**, *6* (13), 3182–3199. https://doi.org/10.1039/c8tc00202a.

(9) Exner, A. T.; Pavlichenko, I.; Baierl, D.; Schmidt, M.; Derondeau, G.; Lotsch, B. V.; Lugli, P.; Scarpa, G. A Step towards the Electrophotonic Nose: Integrating 1D Photonic Crystals with Organic Light-Emitting Diodes and Photodetectors. *Laser Photonics Rev.* **2014**, *8* (5), 726–733. https://doi.org/10.1002/lpor.201300220.

(10) Pavlichenko, I.; Broda, E.; Fukuda, Y.; Szendrei, K.; Hatz, A. K.; Scarpa, G.; Lugli, P.; Bräuchle, C.; Lotsch, B. V. Bringing One-Dimensional Photonic Crystals to a New Light: An Electrophotonic Platform for Chemical Mass Transport Visualisation and Cell Monitoring. *Mater. Horizons* **2015**, *2* (3), 299–308. https://doi.org/10.1039/c4mh00195h.

(11) Liu, J.; Yao, M.; Shen, L. Third Generation Photovoltaic Cells Based on Photonic Crystals. *J. Mater. Chem. C* **2019**, *7* (11), 3121–3145.

(12) Lu, Y.-J.; Chang, C.-H.; Lin, C.-L.; Wu, C.-C.; Hsu, H.-L.; Chen, L.-J.; Lin,

Y.-T.; Nishikawa, R. Achieving Three-Peak White Organic Light-Emitting Devices Using Wavelength-Selective Mirror Electrodes. *Appl. Phys. Lett.* **2008**, *92* (12), 111.

(13) Fan, Y.; Zhang, H.; Chen, J.; Ma, D. Three-Peak Top-Emitting White Organic Emitting Diodes with Wide Color Gamut for Display Application. *Org. Electron.* **2013**, *14* (7), 1898–1902.

(14) Ji, W.; Zhang, L.; Zhang, T.; Liu, G.; Xie, W.; Liu, S.; Zhang, H.; Zhang, L.; Li, B. Top-Emitting White Organic Light-Emitting Devices with a One-Dimensional Metallic-Dielectric Photonic Crystal Anode. *Opt. Lett.* **2009**, *34* (18), 2703–2705.

(15) Dutta Choudhury, S.; Badugu, R.; Lakowicz, J. R. Directing Fluorescence with Plasmonic and Photonic Structures. *Acc. Chem. Res.* **2015**, *48* (8), 2171–2180.

(16) Park, B.; Kim, M.-N.; Kim, S. W.; Park, J. H. Photonic Crystal Film with Three Alternating Layers for Simultaneous R, G, B Multi-Mode Photonic Band-Gaps. *Opt. Express* **2008**, *16* (19), 14524–14531.

(17) Chen, Z. H.; Liang, L.; Wang, Y.; Qiao, N.; Gao, J.; Gan, Z.; Yang, Y. Tunable High Reflective Bands to Improve Quantum Dot White Light-Emitting Diodes. *J. Mater. Chem. C* **2017**, *5* (5), 1149–1154. https://doi.org/10.1039/c6tc05129d.

(18) Wang, X.; Hu, X.; Li, Y.; Jia, W.; Xu, C.; Liu, X.; Zi, J. Enlargement of Omnidirectional Total Reflection Frequency Range in One-Dimensional Photonic Crystals by Using Photonic Heterostructures. *Appl. Phys. Lett.* **2002**, *80* (23), 4291–4293.

(19) Wang, L.; Wang, Z.; Wu, Y.; Chen, L.; Wang, S.; Chen, X.; Lu, W. Enlargement of the Nontransmission Frequency Range of Multiple-Channeled Filters by the Use of Heterostructures. *J. Appl. Phys.* **2004**, *95* (2), 424–426. https://doi.org/10.1063/1.1634367.

(20) Zhang, L.; Wang, J.; Tao, S.; Geng, C.; Yan, Q. Universal Fluorescence Enhancement Substrate Based on Multiple Heterostructure Photonic Crystal with Super-Wide Stopband and Highly Sensitive Cr (VI) Detecting Performance. *Adv. Opt. Mater.* **2018**, *6* (11), 1701344.

(21) Shen, Y.; Ye, D.; Celanovic, I.; Johnson, S. G.; Joannopoulos, J. D.; Soljačić, M. Optical Broadband Angular Selectivity. *Science (80-. ).* **2014**, *343* (6178), 1499–1501.

(22) Wu, Z.; Lee, D.; Rubner, M. F.; Cohen, R. E. Structural Color in Porous, Superhydrophilic, and Self-Cleaning SiO2/TiO2 Bragg Stacks. *Small* **2007**, *3* (8), 1445–1451.

(23) Alfrey Jr, T.; Gurnee, E. F.; Schrenk, W. J. Physical Optics of Iridescent Multilayered Plastic Films. *Polym. Eng. Sci.* **1969**, *9* (6), 400–404.

(24) Kou, D.; Zhang, S.; Lutkenhaus, J. L.; Wang, L.; Tang, B.; Ma, W. Porous Organic/Inorganic Hybrid One-Dimensional Photonic Crystals for Rapid Visual Detection of Organic Solvents. *J. Mater. Chem. C* **2018**, *6* (11), 2704–2711.

(25) Qiu, L.; Liu, Q.; Zhang, R.; Wen, T. Distributed Bragg Reflectors with High Robustness and Responsiveness from UV-Curable Resins. *Polymer (Guildf).* **2021**, *221*, 123604.

(26) Lecaruyer, P.; Maillart, E.; Canva, M.; Rolland, J. Generalization of the Rouard Method to an Absorbing Thin-Film Stack and Application to Surface Plasmon Resonance. *Appl. Opt.* **2006**, *45* (33), 8419–8423.

(27) Wang, L.; Zhang, S.; Lutkenhaus, J. L.; Chu, L.; Tang, B.; Li, S.; Ma, W. All Nanoparticle-Based P (MMA--AA)/TiO 2 One-Dimensional Photonic Crystal Films with Tunable Structural Colors. *J. Mater. Chem. C* **2017**, *5* (32), 8266–8272.

(28) Kou, D.; Ma, W.; Zhang, S.; Lutkenhaus, J. L.; Tang, B. High-Performance and Multifunctional Colorimetric Humidity Sensors Based on Mesoporous Photonic Crystals and Nanogels. *ACS Appl. Mater. Interfaces* **2018**, *10* (48), 41645–41654.

(29) Kou, D.; Zhang, Y.; Zhang, S.; Wu, S.; Ma, W. High-Sensitive and Stable Photonic Crystal Sensors for Visual Detection and Discrimination of Volatile Aromatic Hydrocarbon Vapors. *Chem. Eng. J.* **2019**, *375*, 121987.

(30) Dou, Y.; Han, J.; Wang, T.; Wei, M.; Evans, D. G.; Duan, X. Fabrication of

MMO--TiO$_2$ One-Dimensional Photonic Crystal and Its Application as a Colorimetric Sensor. *J. Mater. Chem.* **2012**, *22* (28), 14001–14007.

# Supporting Information

# High-Intensity and Uniform Red-Green-Blue Triple Reflective Bands Achieved by Rationally-Designed Ultrathin Heterostructure Photonic Crystals

**Authors:**

Lu Qiu[1,2#], Quan-Shan Liu[1,2#], Rui Zhang[1,2*], Tao Wen[1,2*]

**Section 1**

The general Fresnel reflection coefficients can be represented as[1]:

$$r_s = \frac{n_i \cos\theta_i - n_t \cos\theta_t}{n_i \cos\theta_i + n_t \cos\theta_t} \ , \tag{1}$$

$$r_p = \frac{n_i \cos\theta_t - n_t \cos\theta_i}{n_i \cos\theta_t + n_t \cos\theta_i} \ , \tag{2}$$

Here, s and p refer to s-type and p-type polarization respectively, while *n* represents the refractive index. The subscripts *i* and *t* indicate incidence and transmittance respectively. Besides, $\theta_i$ and $\theta_t$ are the angles of incidence and transmittance on a dielectric interface, respectively.

At normal incidence, both $\theta_i$ and $\theta_t$ are 0. It can be seen that equations (1) and (2) can be reduced to the same result:

$$r = \frac{n_i - n_t}{n_i + n_t} \ , \tag{3}$$

In this case, it is no need to label subscripts s and p. By a simple mathematical transformation, r can be represented as:

$$r = \frac{1 - n_t/n_i}{1 + n_t/n_i} \ , \tag{4}$$

Therefore, the Fresnel reflection coefficient is determined by the refractive index ratio of dielectric media.

**Section 2**

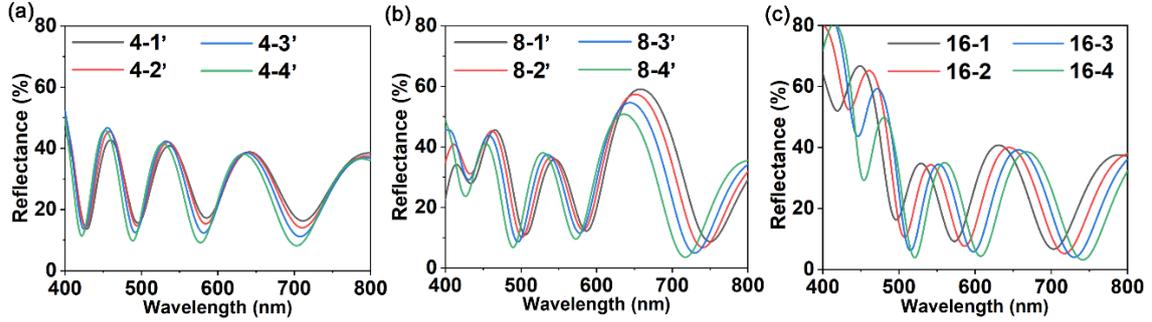

**Figure S1**. (a) The reflection spectra of 4-layer heterostructures with 4 groups of material thicknesses. (b) The reflection spectra of 8-layer heterostructures with 4 groups of material thicknesses. (c) The reflection spectra of 16-layer heterostructures with 4 groups of material thicknesses.

Given that both two categories of stacks hold their partial OPDs exceeding the scope of RGB zones, we designed another 4 sets of material thicknesses for the 4-layer heterostructure, noted as 4-1', 4-2', 4-3', and 4-4' respectively. Regarding precise thickness values, we have 10|90|720|90 for 4-1' (OPD = 3186 nm), 20|75|730|75 for 4-2' (OPD = 3165 nm), 30|60|740|60 for 4-3' (OPD = 3144 nm), and 40|45|750|45 for 4-4' (OPD = 3123 nm), and the corresponding simulated reflection spectra are shown in Figure S1a. Under the same assumption, another 4 groups of material thicknesses can be constructed for the 8-layer heterostructure, represented by 8-1', 8-2', 8-3', and 8-4'. Regarding precise thickness values, we have 10|90|290|90 for 8-1' (OPD = 3276 nm), 20|75|300|75 for 8-2' (OPD = 3234 nm), 30|60|310|60 for 8-3'(OPD = 3192 nm), and 40|45|320|45 for 8-4' (OPD = 3150 nm), and the corresponding simulated reflection spectra are shown in Figure S1b. It can be seen that the three band intensities have an overall worse performance in both cases.

On the other hand, many groups of material thicknesses were tried out for the 16-layer heterostructure, and four of them were picked out, denoted by 16-1, 16-2, 16-3, and 16-4 respectively. Regarding precise thickness values, we have 25|85|45|85 for 16-1 (OPD = 3116 nm), 40|70|60|70 for 16-2(OPD = 3176 nm), 55|55|75|55 for 16-3(OPD = 3236 nm), and 70|40|90|40 for 16-4 (OPD = 3296 nm), and the corresponding

simulated reflection spectra are shown in Figure S1c. Though the data set of material thicknesses in our simulation was limited, good uniformity of the three band intensities had never been made during our attempts.

**Section 3**

By changing material thicknesses, the two partial OPDs held by two different stacks can be tuned easily. With the total OPD keeping fixed at 2885 nm, we set another two groups of material thicknesses for the 12-layer heterostructure, noted by 12-#1 and 12-#2. For 12-#1, we set material thicknesses as 40|75|98|75. Therefore, the magnitude difference of the two partial OPDs is about $2(n_1 d_3 + n_2 d_4) - 2(n_1 d_1 + n_2 d_2) = (585 - 376)\ nm = 209\ nm$. For 12-#2, we set material thicknesses as 60|75|78|75. Therefore, the magnitude difference of the two partial OPDs is about $2(n_1 d_3 + n_2 d_4) - 2(n_1 d_1 + n_2 d_2) = (513 - 448)\ nm = 65\ nm$. For the 12-layer heterostructure mentioned in the main text with a modified CIE coordinate, the magnitude difference is about $2(n_1 d_3 + n_2 d_4) - 2(n_1 d_1 + n_2 d_2) = (549 - 412)\ nm = 137\ nm$.

It should be noted that the two partial OPDs also reveal the stop-band positions (SBPs) of the two constituent 1DPCs in the 12-layer heterostructure, since the two partial OPDs are equal to the two SBPs in magnitude respectively. To graphically present the magnitude difference of the two partial OPDs, we could illustrate the relative distance between the two SBPs by calculating the reflection spectra of the two constituent 1DPCs. For convenience, we represent those 1DPCs by 12-#1-thin and 12-#1-thick for the 12-#1 heterostructure, and 12-#2-thin and 12-#2-thick for the 12-#2 heterostructure. Figure S2a,b shows the simulated reflection spectra of the two 12-layer heterostructures, together with those of their constituent 1DPCs. Also, the corresponding CIE coordinates of the two 12-layer heterostructures were obtained, as shown in Figure S4c. It can be seen that band intensity variation results in a CIE coordinate shift.

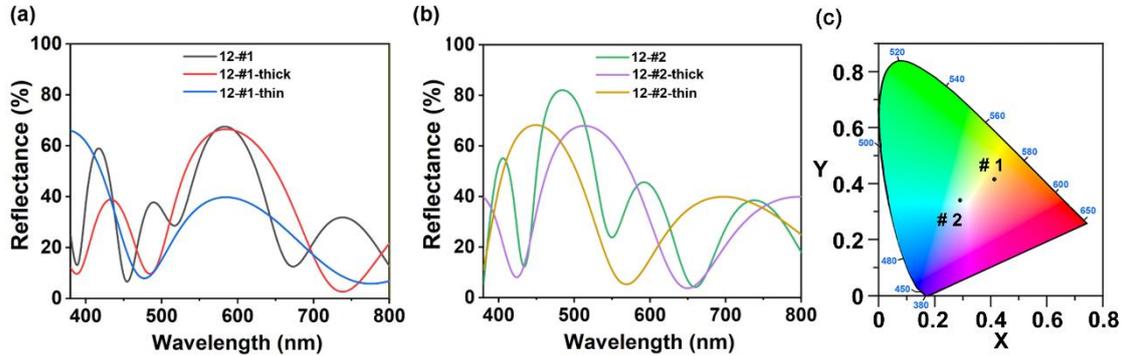

**Figure S2**. (a) The reflection spectrum of the 12-#1 heterostructure and those of the constituent 1DPCs. (b) The reflection spectrum of the 12-#2 heterostructure and those of the constituent 1DPCs. (c) CIE 1931 chromaticity coordinates of the 12-#1 and the 12-#2 heterostructure.

For 12-#1, the distance between the two SBPs is pretty large, and one of the SBP is not located in RGB zones. In the spectrum of the 12-#1 heterostructure, the band intensity at ~481 nm is notably lower than those at ~577 and ~412 nm. This is because the middle reflective band is far from the two SBPs. The case is the opposite for 12-#2, where the distance between the two SBPs is pretty small. In the spectrum of the 12-#2 heterostructure, the band intensity at ~481 nm is notably higher than those at ~ 577 and ~412 nm. The reason is that the two constituent 1DPCs become less distinguishable, causing a traditional 1DPC-like reflection performance. For the 12-layer heterostructure designed in the main text with a modified CIE coordinate, the distance between the two SBPs is desirable. Consequently, the three HRBs of the heterostructure share similar intensities.

**Section 4**

*Materials*

UV resin (HY8208) was provided by Jiangmen Hongye Chemical Co., LTD. Methylammonium bromine (MABr) and lead bromine ($PbBr_2$) were purchased from TCI. 2-bromo-1,1,2-triphenylethylene, 4-aminobenzeneboronic acid hydrochloride, bis(triphenyl-phosphine) palladium(II) diacetate ($Pd(PPh_3)_4$), potassium carbonate

(K$_2$CO$_3$), tetrabutylammonium bromide (TBAB), 2-bromo-2-methylpropionic acid and cyclohexyl isocyanide were purchased from Innochem. *N, N*-dimethylformamide. CdSe and TPE were kindly provided by my friends. Tetrabutyl titanate was purchased from Aladdin Industrial Corporation HCl, deionized water, ethyl acetate, ethanol, chloroform, and toluene were purchased from Sigma-Aldrich and used without further purification.

*Fabrication of heterostructures*

The stock solution of titania sol was prepared according to the previous report.[2] Briefly, 4 mL tetrabutyl titanate was dissolved in 10 mL ethanol in a conical flask. 0.1 mL HCl and 3 mL deionized water were mixed with 4 mL ethanol. Then, this mixture solution was added into the above tetrabutyl titanate solution dropwise in an ice water bath for cooling, and the mixture was stirred for 12 hours at room temperature. The solution of titania sol was diluted with alcohol before use. UV resin was diluted with ethyl acetate before use, and the final thickness of resin layers is controlled by the concentration of resin. Silicon wafers (15 mm ×15 mm) were soaked in acetone for ultrasonic cleaning for 30 mins, and then washed with fresh acetone for 3 times. After cleaning by UV/ozone for 15 min, the silicon wafers were washed using deionized water for several times, and dried with N$_2$ stream.

Briefly, a typical fabrication process of heterostructures is described as follows. The titania sol and the UV resin are alternately spin-coated onto a silicon wafer at 3000 rpm for 30 s. Each resin layer was cured with a UV lamp (6W) for 10 min, whereas each titania layer was baked at 50°C for 1 min after spin-coating. The first layer deposited on the Si substrate was the polymer layer in all our experiments.

*Characterization*

Scanning electron microscopy (SEM) micrographs were taken with a JEOL FESEM 6700F electron microscope. All the samples were sputtered with a thin layer of Pt before imaging. The reflectance spectra from 300 to 1000 nm were measured by a fiber optic spectrometer (Ocean Optics, Maya 2000Pro) with standard mirror optics at the incidence angle of 90°. The determination of layer thicknesses and optical constants were recorded with an ellipsometer (SV-2000) at an angle of 65.5° and wavelength at 400-800 nm. The UV-Vis spectra of single-layer films were recorded using a Shimadzu UV-2450 spectrometer.

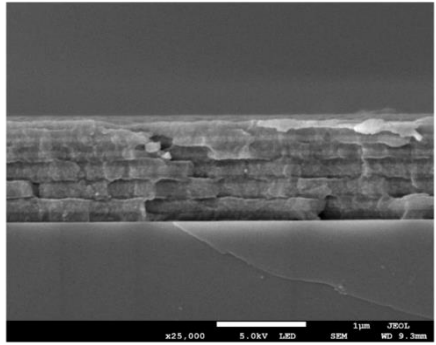

**Figure S3.** The cross-sectional SEM image of the experimental sample. Brighter layers represent $TiO_2$, indicating a higher dielectric constant.

**Section 5**

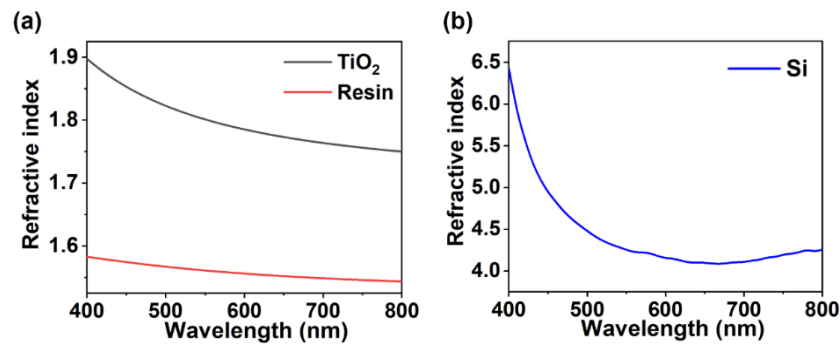

**Figure S4**. (a) The dispersion relationships of the titania sol ($TiO_2$) and the UV-curable resin (resin). (b) The dispersion relationship of the Si substrate.

The refractive indices of the $TiO_2$ and the resin were measured by ellipsometer, while that of the Si substrate was calculated from its reflection spectrum at normal incidence. For all the three experimental materials, the imaginary parts of the refractive indices are relatively low in the visible light region, so they were ignored during our simulation.

**Section 6**

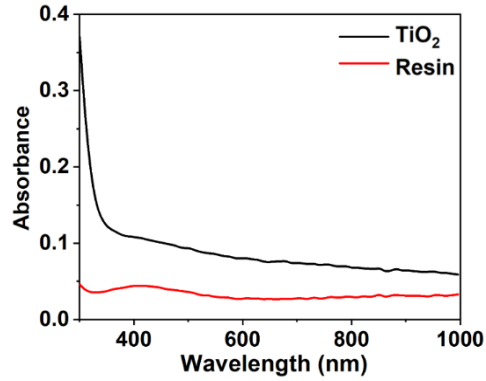

**Figure S5.** The light absorption spectra of the experimental materials.

The light absorption spectra of the experimental materials were measured by a Shimadzu UV-2450 spectrometer. It can be observed that the light absorption losses for both the $TiO_2$ and the resin are relatively low at the visible light range, so they were ignored during our simulation.

**Section 7**

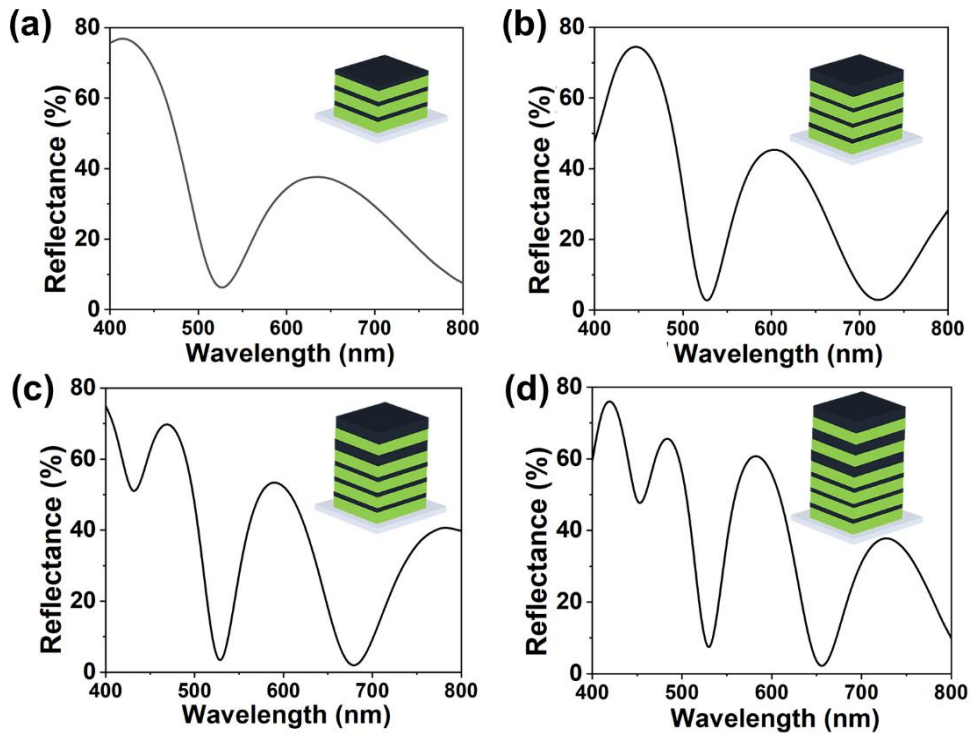

**Figure S6.** The simulated reflection spectra of samples in different stages (a) 0'+3'. (b) 1'+3'. (c) 2'+3'. (d) 3'+3'.

We also calculated the reflection spectra of samples ranging from Air/3-stack thinner 1DPC/Si substrate (short for 0'+3') to Air/3-stack thicker 1DPC/3-stack thinner

1DPC/Si substrate (short for 3'+3'), as shown in Figure S8 a-d.

Reference

(1)  Lecaruyer, P.; Maillart, E.; Canva, M.; Rolland, J. Generalization of the Rouard Method to an Absorbing Thin-Film Stack and Application to Surface Plasmon Resonance. *Appl. Opt.* **2006**, *45* (33), 8419–8423.

(2)  Dastan, D. Effect of Preparation Methods on the Properties of Titania Nanoparticles: Solvothermal versus Sol–Gel. *Appl. Phys. A* **2017**, *123* (11), 699. https://doi.org/10.1007/s00339-017-1309-3.